\documentclass{article}
\usepackage{cite}
\usepackage{graphicx}
\usepackage{dcolumn}
\usepackage{amsmath}

\begin{document}

\date{}
\title{On the spectra of coupled harmonic oscillators}
\author{Francisco M. Fern\'{a}ndez\thanks{%
fernande@quimica.unlp.edu.ar} \\
INIFTA, DQT, Sucursal 4, C.C 16, \\
1900 La Plata, Argentina}
\maketitle

\begin{abstract}
We discuss the diagonalization of a general Hamiltonian operator for a set
of coupled harmonic oscillators and determine the conditions for the
existence of bound states. We consider the particular cases of two and three
oscillators studied previously and show the conditions for bound states in
the latter example that have been omitted in an earlier treatment of this
model.
\end{abstract}

\section{Introduction}

\label{sec:intro}

Models of coupled harmonic oscillators (CHO) have been extensively used to
approximate and illustrate a wide variety of physical problems\cite{MS96}
(and references therein). They appear, for example, in the analysis of small
oscillations in classical mechanics\cite{G80} and in the theory of molecular
vibrations\cite{WDC55}. There has recently been great interest in the
analysis of the symmetries of CHO and the two-mode squeezed states\cite
{HKNY93,HKN95}.The model proved to be a pedagogical illustrative example of
Feynman's rest of the universe\cite{HKN99} and suitable for the study of
entanglement in quantum mechanics\cite{JMM11,M18,MJ20}. The starting point
of these studies consists of rewriting the Hamiltonian in diagonal form by
means of two canonical transformations of the coordinates and their
conjugate momenta\cite{HKNY93,HKN95,HKN99,JMM11,M18,MJ20} but it seems that
the results in some of the papers are not correct\cite{M18}.

The parameters in the Hamiltonians for those CHO should be chosen with care
in order to have bound states. The conditions have been completely specified
in the case of some two-dimensional models\cite{HKNY93,HKN95}, only
partially specified in some cases\cite{JMM11} and omitted in others\cite
{HKN99,M18}. In the only treatment of three CHO the parameter conditions for
bound states were completely ignored, most probably because the second
canonical transformation, based on the $SU(3)$ group, far from solving the
problem leads to six transcendental equations that the authors never solved%
\cite{MJ20}. The one-step algorithm based on the diagonalization of two
symmetric matrices\cite{G80,WDC55} appears to be simpler and more
straightforward than the one just mentioned\cite
{HKNY93,HKN95,HKN99,JMM11,M18,MJ20} but it seems to have been overlooked in
the latter treatments of the CHO. A pedagogical geometrical interpretation
of this one-step algorithm in the case of two oscillators looks rather
confusing because it resorts to more than one transformation\cite{A89}.

The purpose of this paper is the application of the one-step algorithm\cite
{G80,WDC55} to the particular cases of two and three oscillators studied
recently\cite{HKNY93,HKN95,HKN99,JMM11,M18,MJ20} with the purpose of
determining the conditions that the coefficients of the Hamiltonian for the
three-oscillator model\cite{MJ20} should satisfy so that there are bound
states.

In section~\ref{sec:the method} we develop the approach for a
quantum-mechanical CHO model instead of using the results for the
classical version considered earlier\cite{G80,WDC55,A89}. Although
the frequencies of the normal modes of both the classical and
quantum-mechanical CHO are exactly the same it is worth developing
the approach for the latter case because it does not appear to be
so widely discussed\cite{G80,WDC55,A89}. In
section~\ref{sec:examples} we apply the general results to the
particular cases of two and three CHO already mentioned above\cite
{HKNY93,HKN95,HKN99,JMM11,M18,MJ20}. In
section~\ref{sec:conclusions} we summarize the main results and
draw conclusions and at the end of this paper the reader will find
the Appendix~\ref{sec:appendix} with the necessary and sufficient
conditions for bounds states in the cases of four and five coupled
harmonic oscillators.

\section{Diagonalization of the model Hamiltonian}

\label{sec:the method}

We consider a quantum-mechanical system with $N$ coordinates $x_{i}$ and
conjugate momenta $p_{j}$ that satisfy the well known commutation relations $%
[x_{i},p_{j}]=i\hbar \delta _{ij}$, $i,j=1,2,\ldots ,N$. The
Hamiltonian is a quadratic function of these dynamical variables
\begin{equation}
H=\frac{1}{2}\left( \mathbf{p}^{t}\mathbf{Tp}+\mathbf{x}^{t}\mathbf{Vx}%
\right) ,  \label{eq:Hamiltonian}
\end{equation}
where $\mathbf{p}^{t}=\left( p_{1}\;p_{2}\;\ldots \;p_{N}\right) $, $\mathbf{%
x}^{t}=\left( x_{1}\;x_{2}\;\ldots \;x_{N}\right) $ ($t$ stands for
transpose) and $\mathbf{T}$, $\mathbf{V}$ are $N\times N$ real symmetric
matrices.

We carry out the canonical transformation
\begin{equation}
\mathbf{x}=\mathbf{Cx}^{\prime },\;\mathbf{p}=\left( \mathbf{C}^{t}\right)
^{-1}\mathbf{p}^{\prime },  \label{eq:can_transf}
\end{equation}
so that the new momenta $\mathbf{p}^{\prime t}=\left( p_{1}^{\prime
}\;p_{2}^{\prime }\;\ldots \;p_{N}^{\prime }\right) $ and coordinates $%
\mathbf{x}^{\prime t}=\left( x_{1}^{\prime }\;x_{2}^{\prime }\;\ldots
\;x_{N}^{\prime }\right) $ satisfy $[x_{i}^{\prime },p_{j}^{\prime }]=i\hbar
\delta _{ij}$, $i,j=1,2,\ldots ,N$. We choose the $N\times N$ matrix $%
\mathbf{C}$ so that
\begin{equation}
\mathbf{C}^{-1}\mathbf{T}\left( \mathbf{C}^{t}\right) ^{-1}=\mathbf{I},\;%
\mathbf{C}^{t}\mathbf{VC}=\mathbf{\Lambda ,}  \label{eq:C_conditions}
\end{equation}
where $\mathbf{I}$ is the $N\times N$ identity matrix and $\mathbf{\Lambda }$
is a diagonal matrix with elements $\lambda _{i}$, $i=1,2,\ldots ,N$.
Therefore, the Hamiltonian operator (\ref{eq:Hamiltonian}) becomes
\begin{equation}
H=\frac{1}{2}\left( \mathbf{p}^{\prime t}\mathbf{p}^{\prime }+\mathbf{x}%
^{\prime t}\mathbf{\Lambda x}^{\prime }\right) .  \label{eq:H_diag}
\end{equation}
Clearly there will be bound states provided that $\lambda _{i}>0$, $%
i=1,2,\ldots ,N$ . Because of the commutation relations between the new
coordinates and momenta the eigenvalues are given by
\begin{equation}
E_{\{n\}}=\hbar \sum_{i=1}^{N}\sqrt{\lambda _{i}}\left( n_{i}+\frac{1}{2}%
\right) ,\;\{n\}=\{n_{1},n_{2},\ldots ,n_{N}\},\;n_{i}=0,1,\ldots .
\label{eq:eigenvalues}
\end{equation}
It follows from equations (\ref{eq:C_conditions}) that
\begin{equation}
\mathbf{C}^{-1}\mathbf{TVC}=\mathbf{\Lambda },  \label{eq:diagonalization_TV}
\end{equation}
so that the whole problem reduces to the diagonalization of the
non-symmetric matrix $\mathbf{A}=\mathbf{TV}$. This result is well known in
molecular spectroscopy where it has proved suitable for the study of
molecular vibrations in terms of generalized coordinates, although it was
derived in the realm of classical mechanics\cite{WDC55}. A slightly
different, though entirely equivalent, equation has also been derived in the
study of small oscillations in classical mechanics\cite{G80}.

There are alternative ways of obtaining $H$ in diagonal form. If we prefer
diagonalizing symmetric matrices we can define $\mathbf{C}=\mathbf{T}^{1/2}%
\mathbf{U}$ provided $\mathbf{T}$ is positive definite. In this case
equation (\ref{eq:diagonalization_TV}) becomes
\begin{equation}
\mathbf{U}^{-1}\mathbf{T}^{1/2}\mathbf{VT}^{1/2}\mathbf{U}=\mathbf{\Lambda }.
\label{eq:diagonalization_sym_mat}
\end{equation}
Since $\mathbf{S}=\mathbf{T}^{1/2}\mathbf{VT}^{1/2}$ is symmetric then $%
\mathbf{U}$ is orthogonal ($\mathbf{U}^{-1}=\mathbf{U}^{t}$) and we can use
well known efficient diagonalization routines. The calculation of $\mathbf{T}%
^{1/2}$ is particularly straightforward when $\mathbf{T}$ is diagonal (as in
the examples mentioned above\cite{HKNY93,HKN95,HKN99,JMM11,M18,MJ20}). Any $%
N\times N$ orthogonal matrix has only $N(N-1)/2$ independent matrix
elements. Therefore, for $N=2$ and $N=3$ we can write $\mathbf{U}$ in terms
of two and tree independent quantities (angles, for example), respectively%
\cite{HKNY93,HKN95,HKN99,JMM11,M18,MJ20}.

Notice that $x_{i}^{\prime }$ and $p_{i}^{\prime }$ do not longer have units
of length and momentum, respectively, because $\mathbf{C}$ has units of $%
mass^{-1/2}$ (assuming that $\mathbf{T}$ has units of $mass^{-1}$). However,
we obtain the correct eigenvalues because the transformed dynamical
variables satisfy the standard canonical commutation relations. But if we
want the dynamical variables to keep their standard physical units we simply
change the conditions (\ref{eq:C_conditions}) to
\begin{equation}
\mathbf{C}^{-1}\mathbf{T}\left( \mathbf{C}^{t}\right) ^{-1}=\frac{1}{m}%
\mathbf{I},\;\mathbf{C}^{t}\mathbf{VC}=\mathbf{K},  \label{eq:C_conditions_2}
\end{equation}
where $m$ is an arbitrary mass and $\mathbf{K}$ a diagonal matrix. In this
case $\mathbf{C}$ is dimensionless, the diagonalization equation becomes
\begin{equation}
\mathbf{C}^{-1}\mathbf{TVC}=\frac{1}{m}\mathbf{K}=\mathbf{\Lambda },
\label{eq:diagonalization_TV_dim}
\end{equation}
and the resulting Hamiltonian reads
\begin{equation}
H=\frac{1}{2m}\mathbf{p}^{\prime t}\mathbf{p}^{\prime }+\frac{1}{2}\mathbf{x}%
^{\prime t}\mathbf{Kx}^{\prime }.  \label{eq:H_diag_2}
\end{equation}
It is clear that its eigenvalues are exactly those given above in equation (%
\ref{eq:eigenvalues}) and, consequently, independent of the arbitrary mass $%
m $. This fact may appear to be strange at first sight but one has to take
into consideration that the transformation $\mathbf{C}^{-1}\mathbf{T}\left(
\mathbf{C}^{t}\right) ^{-1}$ is merely a normalization condition for the
eigenvectors of $\mathbf{A}$ that are the columns of the matrix $\mathbf{C}$%
. If one feels uncomfortable about having an arbitrary mass in the
intermediate equations one may set it to be, for example, the geometric mean
$m=\left( m_{1}m_{2}\ldots m_{N}\right) ^{1/N}$ (when $T_{ij}=\delta
_{ij}/m_{i}$, $i,j=1,2,\ldots ,N$) as in earlier studies of the particular
cases $N=2$\cite{HKNY93,HKN95,HKN99,JMM11,M18} and $N=3$\cite{MJ20}.

The symmetric matrix $\mathbf{S}$ is particularly useful for determining the
values of the model parameters that are compatible with positive eigenvalues
$\lambda _{i}$ and, consequently, bound-state solutions. It is well known
that \textit{a symmetric matrix is positive definite if and only if each of
its leading principal minors is positive}\cite{M00}. This theorem will prove
useful in the analysis of the examples below.

\section{Examples}

\label{sec:examples}

We first consider the particular case of $N=2$ coupled harmonic oscillators%
\cite{HKNY93,HKN95,HKN99,JMM11,M18}
\begin{equation}
H=\frac{1}{2m_{1}}p_{1}^{2}+\frac{1}{2m_{2}}p_{2}^{2}+\frac{1}{2}\left(
C_{1}x_{1}^{2}+C_{2}x_{2}^{2}+C_{3}x_{1}x_{2}\right) .  \label{eq:H_N=2}
\end{equation}
In this case $\mathbf{T}$ is positive-definite and diagonal which renders
the calculation of $T^{1/2}$ trivial.

The matrices
\begin{eqnarray}
\mathbf{A} &=&\left(
\begin{array}{ll}
\frac{C_{1}}{m_{1}} & \frac{C_{3}}{2m_{1}} \\
\frac{C_{3}}{2m_{2}} & \frac{C_{2}}{m_{2}}
\end{array}
\right) ,  \nonumber \\
\mathbf{S} &=&\left(
\begin{array}{ll}
\frac{C_{1}}{m_{1}} & \frac{C_{3}}{2\sqrt{m_{1}m_{2}}} \\
\frac{C_{3}}{2\sqrt{m_{1}m_{2}}} & \frac{C_{2}}{m_{2}}
\end{array}
\right) ,  \label{eq:A,S_N=2}
\end{eqnarray}
have the characteristic polynomial
\begin{equation}
\lambda ^{2}-\frac{\lambda \left( m_{2}C_{1}+m_{1}C_{2}\right) }{m_{1}m_{2}}+%
\frac{4C_{1}C_{2}-C_{3}^{2}}{4m_{1}m_{2}}=0,  \label{eq:charpoly_N=2}
\end{equation}
which will have two real and positive roots provided that
\begin{equation}
m_{2}C_{1}+m_{1}C_{2}>0,\;4C_{1}C_{2}-C_{3}^{2}>0.
\label{eq:eigenv_cond_N=2}
\end{equation}
It follows from these two conditions that $C_{1},C_{2}>0$, already mentioned
in some treatments of this model\cite{HKNY93,HKN95}. Notice that it is only
necessary to specify two conditions instead of three and that some of the
conditions are omitted in some earlier treatments of this model\cite
{JMM11,HKN99,M18}. The two principal minors of $\mathbf{S}$ are positive
provided that $C_{1}>0$ and $4C_{1}C_{2}-C_{3}^{2}>0$ which are the
necessary and sufficient conditions for positive definiteness and,
consequently, positive eigenvalues $\lambda _{i}$. They are equivalent to
those discussed above.

The eigenvalues of $\mathbf{A}$ and $\mathbf{S}$ are
\begin{eqnarray}
\lambda _{1} &=&\frac{m_{1}C_{2}+m_{2}C_{1}-R}{2m_{1}m_{2}},\;\lambda _{2}=%
\frac{m_{1}C_{2}+m_{2}C_{1}+R}{2m_{1}m_{2}},  \nonumber \\
R &=&\sqrt{\left( m_{2}C_{1}-m_{1}C_{2}\right) ^{2}+m_{1}m_{2}C_{3}^{2}}.
\label{eq:eigen_N=2}
\end{eqnarray}

The second particular example is given by the three coupled oscillators\cite
{MJ20}
\begin{eqnarray}
H &=&\frac{1}{2m_{1}}p_{1}^{2}+\frac{1}{2m_{2}}p_{2}^{2}+\frac{1}{2m_{3}}%
p_{3}^{2}  \nonumber \\
&&+\frac{1}{2}\left( m_{1}\omega _{1}^{2}x_{1}^{2}+m_{2}\omega
_{2}^{2}x_{2}^{2}+m_{3}\omega
_{3}^{2}x_{3}^{2}+D_{12}x_{1}x_{2}+D_{13}x_{1}x_{3}+D_{23}x_{2}x_{3}\right) .
\nonumber \\
&&  \label{eq:H_N=3}
\end{eqnarray}
In this case the matrix $\mathbf{T}$ is also positive-definite and diagonal.
The matrices that are relevant for the diagonalization of this Hamiltonian
operator are
\begin{eqnarray}
\mathbf{A} &=&\left(
\begin{array}{lll}
\omega _{1}^{2} & \frac{D_{12}}{2m_{1}} & \frac{D_{13}}{2m_{1}} \\
\frac{D_{12}}{2m_{2}} & \omega _{2}^{2} & \frac{D_{23}}{2m_{2}} \\
\frac{D_{13}}{2m_{3}} & \frac{D_{23}}{2m_{3}} & \omega _{3}^{2}
\end{array}
\right) ,  \nonumber \\
\mathbf{S} &=&\left(
\begin{array}{lll}
\omega _{1}^{2} & \frac{D_{12}}{2\sqrt{m_{1}m_{2}}} & \frac{D_{13}}{2\sqrt{%
m_{1}m_{3}}} \\
\frac{D_{12}}{2\sqrt{m_{1}m_{2}}} & \omega _{2}^{2} & \frac{D_{23}}{2\sqrt{%
m_{2}m_{3}}} \\
\frac{D_{13}}{2\sqrt{m_{1}m_{3}}} & \frac{D_{23}}{2\sqrt{m_{2}m_{3}}} &
\omega _{3}^{2}
\end{array}
\right) ,  \label{eq:Matrices_N=3}
\end{eqnarray}
where the symmetric matrix $\mathbf{S}$ is identical to the matrix $\mathbf{R%
}$ derived by Merdaci and Jellal\cite{MJ20}. These authors claimed to have
solved this problem exactly but they merely derived six transcendental
equations for the six independent elements of their matrix $\mathbf{R}$ in
terms of its three eigenvalues $\Sigma _{i}^{2}$ and three angles that
define the matrix elements of the transformation matrix $\mathbf{M}$
(identical to present orthogonal matrix $\mathbf{U}$).

The characteristic polynomial of any of those matrices (multiplied by $-1$)
is
\begin{eqnarray}
\lambda ^{3} &-&a\lambda ^{2}+b\lambda -c=0,  \nonumber \\
a &=&\left( \omega _{1}^{2}+\omega _{2}^{2}+\omega _{3}^{2}\right) ,
\nonumber \\
b &=&\omega _{1}^{2}\omega _{2}^{2}+\omega _{1}^{2}\omega _{3}^{2}+\omega
_{2}^{2}\omega _{3}^{2}-\frac{D_{12}^{2}}{4m_{1}m_{2}}-\frac{D_{13}^{2}}{%
4m_{1}m_{3}}-\frac{D_{23}{}^{2}}{4m_{2}m_{3}},  \nonumber \\
c &=&\omega _{1}^{2}\omega _{2}^{2}\omega _{3}^{2}-\left( \frac{%
D_{12}^{2}\omega _{3}^{2}}{4m_{1}m_{2}}+\frac{D_{13}^{2}\omega _{2}^{2}}{%
4m_{1}m_{3}}+\frac{D_{23}{}^{2}\omega _{1}^{2}}{4m_{2}m_{3}}-\frac{%
D_{12}D_{13}D_{23}}{4m_{1}m_{2}m_{3}}\right) .  \label{eq:charpoly_N=3}
\end{eqnarray}
If the three roots are real and positive, then $b>0$ and $c>0$. These
conditions are necessary but not sufficient because they are also compatible
with one positive root and two complex-conjugate ones with positive real
part. In order to remove the latter possibility we add the discriminant\cite
{BPR03} of the characteristic polynomial
\begin{equation}
\Delta =\left( \lambda _{1}-\lambda _{2}\right) ^{2}\left( \lambda
_{1}-\lambda _{3}\right) ^{2}\left( \lambda _{2}-\lambda _{3}\right)
^{2}=a^{2}b^{2}-4a^{3}c+18abc-4b^{3}-27c^{2}\geq 0  \label{eq:dscriminant}
\end{equation}
Merdaci and Jellal\cite{MJ20} did not derive any conditions for bound states
probably because they did not solve their six transcendental equations which
are too complicated for such an analysis.

We can derive two remarkably simpler necessary and sufficient conditions for
the existence of bound states from two of the three leading principal minors
of the matrix $\mathbf{S}$:
\begin{eqnarray}
4m_{1}m_{2}\omega _{1}^{2}\omega _{2}^{2}-D_{12}^{2} &>&0,  \nonumber \\
4m_{1}m_{2}m_{3}\omega _{1}^{2}\omega _{2}^{2}\omega
_{3}^{2}+D_{12}D_{13}D_{23}-m_{1}\omega _{1}^{2}D_{23}^{2}-m_{2}\omega
_{2}^{2}D_{13}^{2}-m_{3}\omega _{3}^{2}D_{12}^{2} &>&0.  \nonumber \\
&&  \label{eq:conditions_N=3}
\end{eqnarray}
Notice that one of the conditions has been omitted because it is trivial in
this case ($\omega _{1}^{2}>0$). It is worth mentioning that all the results
about entanglement discussed by Merdaci and Jellal\cite{MJ20} are not valid
unless the model parameters satisfy the two conditions (\ref
{eq:conditions_N=3}).

Merdaci and Jellal\cite{MJ20} tested their unsolved equations by uncoupling
one of the oscillators and restricting the problem to just two coupled
oscillators. This particular case can be achieved by choosing $%
D_{13}=D_{23}=0$. If we do exactly the same we recover the results of two
coupled oscillators discussed above (plus, of course an eigenvalue $\lambda
_{3}=\omega _{3}^{2}$ coming from the uncoupled oscillator). The two
conditions for bound states  (\ref{eq:conditions_N=3}) reduce to just the
first one.

The analytical expressions for the eigenvalues $\lambda _{i}$ and the
transformation matrix $\mathbf{C}$ are quite cumbersome in the general case
(probably the reason why Merdaci and Jellal\cite{MJ20} did not attempt to
solve their equations (12-17)). However, the particular case of three
identical oscillators is remarkably simple and most useful for testing the
general theoretical results given above.

If we set $m_{i}=m$, $\omega _{i}=\omega $, $D_{ij}=D$, $i,j=1,2,3$, we have
\begin{equation}
\mathbf{A}=\mathbf{S}=\frac{1}{2m}\left(
\begin{array}{lll}
m\omega ^{2} & D & D \\
D & m\omega ^{2} & D \\
D & D & m\omega ^{2}
\end{array}
\right) ,
\end{equation}
with eigenvalues
\begin{equation}
\lambda _{1}=\lambda _{2}=\omega ^{2}-\frac{D}{2m},\;\lambda _{3}=\omega
^{2}+\frac{D}{m}.
\end{equation}
This problem is particularly simple because
$\mathbf{TV}=\mathbf{VT}$ which explains why
$\mathbf{A}=\mathbf{S}$. From the eigenvalues we conclude that
there are bound states only when $-m\omega ^{2}<D<2m\omega ^{2}$.
On the other hand, from the three leading principal minors we
obtain $m\omega ^{2}>0 $ (trivial) and
\begin{equation}
4m^{2}\omega ^{4}-D^{2}>0,\;m\omega ^{2}+D>0,
\end{equation}
that lead to exactly the same conditions derived from the eigenvalues.

The calculation of the eigenvectors of the matrix $\mathbf{S}$ is also
extremely simple and we obtain the transformation matrix
\begin{equation}
\mathbf{C}=\mathbf{U}=\frac{1}{\sqrt{6}}\left(
\begin{array}{lll}
\sqrt{3} & 1 & \sqrt{2} \\
0 & -2 & \sqrt{2} \\
-\sqrt{3} & 1 & \sqrt{2}
\end{array}
\right)
\end{equation}
so that $\mathbf{x}=\mathbf{Ux}^{\prime }$ and $\mathbf{p}=\mathbf{Up}%
^{\prime }$.

\section{Conclusions}

\label{sec:conclusions}

In order to transform a general Hamiltonian for a set of coupled oscillators
(\ref{eq:Hamiltonian}) into a diagonal form it is only necessary to obtain
the eigenvalues and eigenvectors of either the nonsymmetric matrix $\mathbf{A%
}$ or the symmetric matrix $\mathbf{S}$ as shown in equations (\ref
{eq:diagonalization_TV}) and (\ref{eq:diagonalization_sym_mat}),
respectively. This procedure is more general than the one based on two
canonical transformations that is suitable for the particular case of a
diagonal matrix $\mathbf{T}$\cite{HKNY93,HKN95,HKN99,JMM11,M18,MJ20}.
Besides, the application of the algebraic method proposed by Merdaci and
Jellal\cite{MJ20} appears to become increasingly cumbersome as $N$ increases
(they were unable to solve the resulting equations even for the second
simplest case $N=3$). On the other hand, the expressions shown in section~%
\ref{sec:the method} are valid for all $N$. Notice that it was quite easy to
obtain the necessary and sufficient conditions for the existence of bound
states in the two simplest cases $N=2$ and $N=3$, the latter of which have
not been taken into account before\cite{MJ20}. Besides, it has been argued
that the parameters of the resulting diagonal Hamiltonian operator have not
been derived correctly even in the simplest case $N=2$\cite{M18}. The
approach sketched here (known since long ago for the classical model\cite
{G80,WDC55}) can be straightforwardly applied to a wider variety of
oscillators with more general couplings than those based on a diagonal
matrix $\mathbf{T}$. In particular, the analysis of the matrix $\mathbf{S}$
in terms of its principal minors is one of the simplest ways of determining
the conditions for bound states.

\appendix

\numberwithin{equation}{section}

\section{Necessary and sufficient conditions for bound states in the cases $%
N=4$ and $N=5$}

\label{sec:appendix}

In the case of $N=4$ we should add

\begin{eqnarray}
&&D_{12}^{2}D_{34}^{2}-4D_{12}^{2}m_{3}m_{4}\omega _{3}^{2}\omega
_{4}^{2}+4D_{12}D_{13}D_{23}m_{4}\omega _{4}^{2}-2D_{12}D_{13}D_{24}D_{34}
\nonumber \\
&&-2D_{12}D_{14}D_{23}D_{34}+4D_{12}D_{14}D_{2{4}}{m_{3}}{\omega _{3}}^{2}
\nonumber \\
&&+D_{13}^{2}D_{24}^{2}-4D_{13}^{2}m_{2}m_{4}\omega _{2}^{2}\omega
_{4}^{2}-2D_{13}D_{14}D_{23}D_{24}+4D_{13}D_{14}D_{34}{m_{2}}{\omega _{2}}%
^{2}  \nonumber \\
&&+D_{14}^{2}D_{23}^{2}-4D_{14}^{2}m_{2}m_{3}\omega _{2}^{2}\omega
_{3}^{2}-4D_{23}^{2}m_{1}m_{4}\omega _{1}^{2}{\omega _{4}}^{2}  \nonumber \\
&&+4D_{23}D_{24}D_{34}m_{1}\omega _{1}^{2}-4D_{24}^{2}m_{1}m_{3}\omega
_{1}^{2}\omega _{3}^{2}-4D_{34}^{2}m_{1}m_{2}\omega _{1}^{2}\omega {2}^{2}
\nonumber \\
&&+16m_{1}m_{2}m_{3}m_{4}\omega _{1}^{2}\omega _{2}^{2}\omega _{3}^{2}\omega
_{4}^{2}>0,
\end{eqnarray}
to the two conditions shown above for $N=3$. For $N=5$ we also have
\begin{eqnarray}
&&D_{12}^{2}D_{34}^{2}m_{5}\omega
_{5}^{2}-D_{12}^{2}D_{34}D_{35}D_{45}+D_{12}^{2}D_{35}^{2}m_{4}\omega
_{4}^{2}  \nonumber \\
&&+D_{12}^{2}D_{45}^{2}m_{3}\omega _{3}^{2}-4D_{12}^{2}m_{3}m_{4}m_{5}\omega
_{3}^{2}\omega _{4}^{2}\omega _{5}^{2}  \nonumber \\
&&-D_{12}D_{13}D_{23}D_{45}^{2}+4D_{12}D_{13}D_{23}m_{4}m_{5}\omega
_{4}^{2}\omega _{5}^{2}  \nonumber \\
&&-2D_{12}D_{13}D_{24}D_{34}m_{5}\omega
_{5}^{2}+D_{12}D_{13}D_{24}D_{35}D_{45}  \nonumber \\
&&+D_{12}D_{13}D_{25}D_{34}D_{45}-2D_{12}D_{13}D_{25}D_{35}m_{4}\omega
_{4}^{2}  \nonumber \\
&&-2D_{12}D_{14}D_{23}D_{34}m_{5}\omega
_{5}^{2}+D_{12}D_{14}D_{23}D_{35}D_{45}  \nonumber \\
&&-D_{12}D_{14}D_{24}D_{35}^{2}+4D_{12}D_{14}D_{24}m_{3}m_{5}\omega
_{3}^{2}\omega _{5}^{2}+D_{12}D_{14}D_{25}D_{34}D_{35}  \nonumber \\
&&-2D_{12}D_{14}D_{25}D_{45}m_{3}\omega
_{3}^{2}+D_{12}D_{15}D_{23}D_{34}D_{45}-2D_{12}D_{15}D_{23}D_{35}m_{4}\omega
_{4}^{2}  \nonumber \\
&&+D_{12}D_{15}D_{24}D_{34}D_{35}-2D_{12}D_{15}D_{24}D_{45}m_{3}\omega
_{3}^{2}  \nonumber \\
&&-D_{12}D_{15}D_{25}D_{34}^{2}+4D_{12}D_{15}D_{25}m_{3}m_{4}\omega
_{3}^{2}\omega _{4}^{2}  \nonumber \\
&&+D_{13}^{2}D_{24}^{2}m_{5}\omega
_{5}^{2}-D_{13}^{2}D_{24}D_{25}D_{45}+D_{13}^{2}D_{25}^{2}m_{4}\omega
_{4}^{2}  \nonumber \\
&&+D_{13}^{2}D_{45}^{2}m_{2}\omega _{2}^{2}-4D_{13}^{2}m_{2}m_{4}m_{5}\omega
_{2}^{2}\omega _{4}^{2}\omega _{5}^{2}-2D_{13}D_{14}D_{23}D_{24}m_{5}\omega
_{5}^{2}  \nonumber \\
&&+D_{13}D_{14}D_{23}D_{25}D_{45}+D_{13}D_{14}D_{24}D_{25}D_{35}-D_{13}D_{14}D_{25}^{2}D_{34}
\nonumber \\
&&+4D_{13}D_{14}D_{34}m_{2}m_{5}\omega _{2}^{2}\omega
_{5}^{2}-2D_{13}D_{14}D_{35}D_{45}m_{2}\omega _{2}^{2}  \nonumber \\
&&+D_{13}D_{15}D_{23}D_{24}D_{45}-2D_{13}D_{15}D_{23}D_{25}m_{4}\omega
_{4}^{2}-D_{13}D_{15}D_{24}^{2}D_{35}  \nonumber \\
&&+D_{13}D_{15}D_{24}D_{25}D_{34}-2D_{13}D_{15}D_{34}D_{45}m_{2}\omega
_{2}^{2}+4D_{13}D_{15}D_{35}m_{2}m_{4}\omega _{2}^{2}\omega _{4}^{2}
\nonumber \\
&&+D_{14}^{2}D_{23}^{2}m_{5}\omega
_{5}^{2}-D_{14}^{2}D_{23}D_{25}D_{35}+D_{14}^{2}D_{25}^{2}m_{3}\omega
_{3}^{2}+D_{14}^{2}D_{35}^{2}m_{2}\omega _{2}^{2}  \nonumber \\
&&-4D_{14}^{2}m_{2}m_{3}m_{5}\omega _{2}^{2}\omega _{3}^{2}\omega
_{5}^{2}-D_{14}D_{15}D_{23}^{2}D_{45}+D_{14}D_{15}D_{23}D_{24}D_{35}
\nonumber \\
&&+D_{14}D_{15}D_{23}D_{25}D_{34}-2D_{14}D_{15}D_{24}D_{25}m_{3}\omega
_{3}^{2}-2D_{14}D_{15}D_{34}D_{35}m_{2}\omega _{2}^{2}  \nonumber \\
&&+4D_{14}D_{15}D_{45}m_{2}m_{3}\omega _{2}^{2}\omega
_{3}^{2}+D_{15}^{2}D_{23}^{2}m_{4}\omega
_{4}^{2}-D_{15}^{2}D_{23}D_{24}D_{34}  \nonumber \\
&&+D_{15}^{2}D_{24}^{2}m_{3}\omega _{3}^{2}+D_{15}^{2}D_{34}^{2}m_{2}\omega
_{2}^{2}-4D_{15}^{2}m_{2}m_{3}m_{4}\omega _{2}^{2}\omega _{3}^{2}\omega
_{4}^{2}  \nonumber \\
&&+D_{23}^{2}D_{45}^{2}m_{1}\omega _{1}^{2}-4D_{23}^{2}m_{1}m_{4}m_{5}\omega
_{1}^{2}\omega _{4}^{2}\omega _{5}^{2}  \nonumber \\
&&+4D_{23}D_{24}D_{34}m_{1}m_{5}\omega _{1}^{2}\omega
_{5}^{2}-2D_{23}D_{24}D_{35}D_{45}m_{1}\omega
_{1}^{2}-2D_{23}D_{25}D_{34}D_{45}m_{1}\omega _{1}^{2}  \nonumber \\
&&+4D_{23}D_{25}D_{35}m_{1}m_{4}\omega _{1}^{2}\omega
_{4}^{2}+D_{24}^{2}D_{35}^{2}m_{1}\omega _{1}^{2}  \nonumber \\
&&-4D_{24}^{2}m_{1}m_{3}m_{5}\omega _{1}^{2}\omega _{3}^{2}\omega
_{5}^{2}-2D_{24}D_{25}D_{34}D_{35}m_{1}\omega _{1}^{2}  \nonumber \\
&&+4D_{24}D_{25}D_{45}m_{1}m_{3}\omega _{1}^{2}\omega
_{3}^{2}+D_{25}^{2}D_{34}^{2}m_{1}\omega _{1}^{2}  \nonumber \\
&&-4D_{25}^{2}m_{1}m_{3}m_{4}\omega _{1}^{2}\omega _{3}^{2}\omega
_{4}^{2}-4D_{34}^{2}m_{1}m_{2}m_{5}\omega _{1}^{2}\omega _{2}^{2}\omega
_{5}^{2}  \nonumber \\
&&+4D_{34}D_{35}D_{45}m_{1}m_{2}\omega _{1}^{2}\omega
_{2}^{2}-4D_{35}^{2}m_{1}m_{2}m_{4}\omega _{1}^{2}\omega _{2}^{2}\omega
_{4}^{2}  \nonumber \\
&&-4D_{45}^{2}m_{1}m_{2}m_{3}\omega _{1}^{2}\omega _{2}^{2}\omega
_{3}^{2}+16m_{1}m_{2}m_{3}m_{4}m_{5}\omega _{1}^{2}\omega _{2}^{2}\omega
_{3}^{2}\omega _{4}^{2}\omega _{5}^{2}>0
\end{eqnarray}
in addition to the three conditions indicated above.

\end{document}